\documentclass[twocolumn,english,aps,prl,10pt,superscriptaddress,showpacs]{revtex4}
\usepackage[T1]{fontenc}
\usepackage{amsmath,graphicx,amssymb,epsfig,babel,dsfont}

\begin{document}

\title{Thermal memristor and neuromorphic networks for manipulating heat flow}

\author{Philippe Ben-Abdallah}
\email{pba@institutoptique.fr} 
\affiliation{Laboratoire Charles Fabry, UMR 8501, Institut d'Optique, CNRS, Universit\'{e} Paris-Saclay,
2 Avenue Augustin Fresnel, 91127 Palaiseau Cedex, France}
\affiliation{Universit\'{e} de Sherbrooke, Department of Mechanical Engineering, Sherbrooke, PQ J1K 2R1, Canada}



\begin{abstract}
A memristor is one of four fundamental two-terminal solid elements in electronics. In addition with the resistor, the capacitor and the inductor, this passive element relates the electric charges to current in solid state elements. Here we report the existence of a thermal analog for this element made with metal-insulator transition materials. We demonstrate that these memristive systems can be used to create thermal neurons opening so the way to neuromophic networks for smart thermal management and  information treatment. 
\end{abstract}

\maketitle

During almost two centuries it was admitted that only three fundamental passive elements, the resistor, the capacitor and the inductor were the building blocks to relate voltage $v$, current $i$, charge $q$, and magnetic flux $\varphi$ in solid elements. However, in 1971 Chua~\cite{Chua} envisioned, through symmetry arguments, the existence of another fundamental element, the memristor a two-terminal non-linear component relating electric charge to flux in electronics circuits. In 2008 Strukov et al.~\cite{Strukov} shown using tunable doped metal-oxide-semiconductors films that this vision was true. The basic mathematical modelling of a memristive system typically takes the following form
\begin{equation}
\begin{split}
v=R(x,w)i,\,
\\ \frac{dx}{dt}=f(x,w), \label{Eq:Conductance}
\end{split}
\end{equation}
where $x$ is a state variable and $R$ is a generalized resistance which depends on this variable and on either the voltage (i.e. $w=v$ for a voltage-controlled memristor) or on the intensity (i.e. $w=i$ for a current-controlled memristor). The distinction between memristive systems and arbitrary dynamical systems is the fact that the voltage  $v$ (output) is always zero when the current $i$ (input) is zero, resulting in zero-crossing Lissajous $v- i$ curves. In this Letter we extend this concept to the heat transport by conduction and we explore the possibilities offered by thermal memristive systems to manage heat exchanges and make information treatment with heat rather than with electric currents as suggested by Li et al.~\cite{BaowenLiEtAl2012}. 

\begin{figure}
\includegraphics[angle=0,scale=0.35]{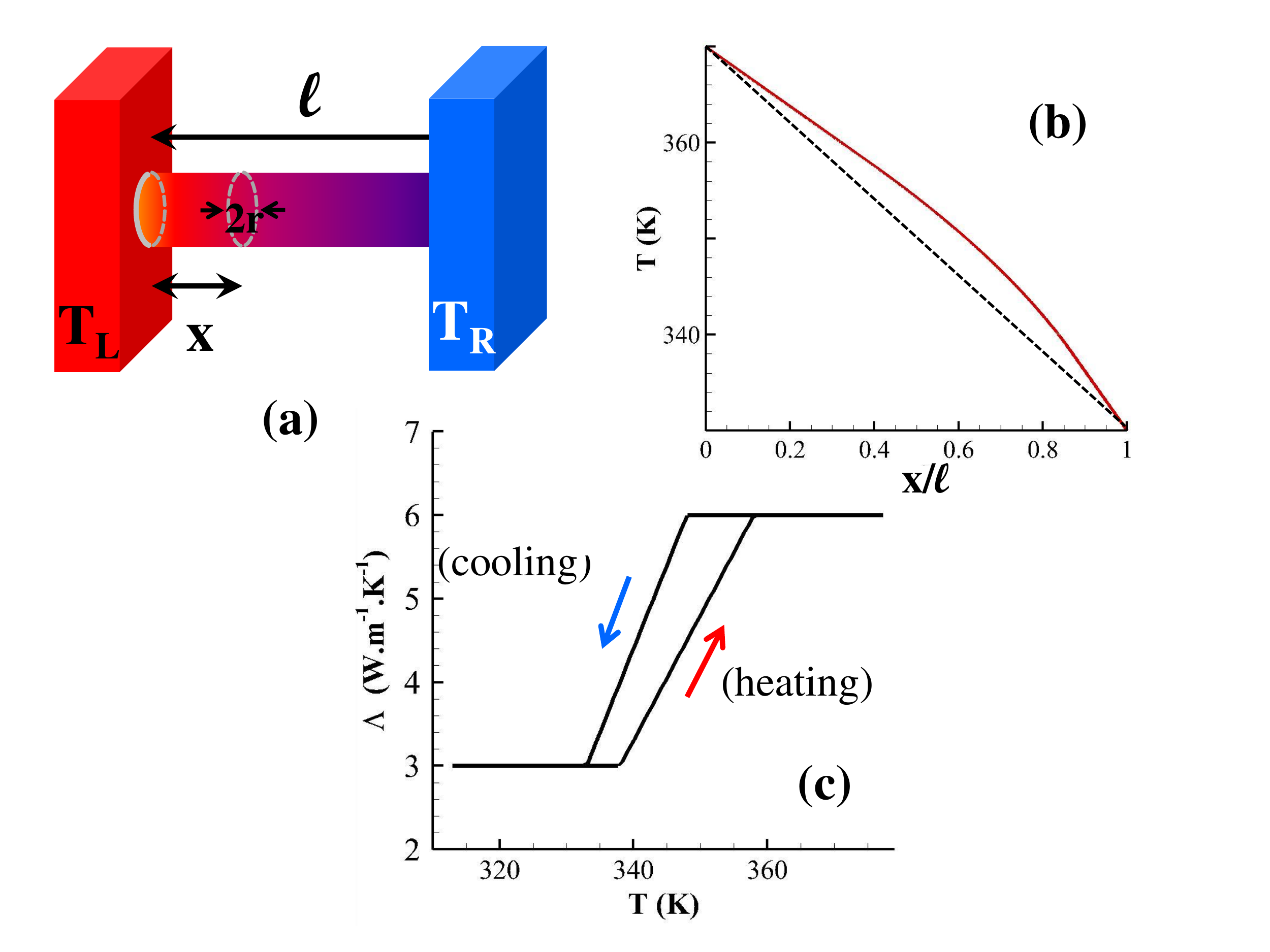}
\caption{(a) Sketch of a phase-change thermal memristor. The wire is a made with a wire/coated wire made with a metal-insulator material in contact with a hot source at temperature $T_L$ and a sink at temperature $T_R$. The position $x$ of phase-change front evoles with respct to $T_L$ and $T_R$ giving rise to  (b) a non-linear temperature profilealong the wire. The linear profile correspond to the temperature in diffusive regime. (c) Thermal conductivity of MIT material  through the phase transition~\cite{Oh} for both heating and cooling processes.}
\label{Fig_1}
\end{figure}

The basic system we consider is sketched on Fig. 1-a. It is a cylindrical wire of radius $r$ and length $l>>r$ made with vanadium dioxide ($VO_2$) a metal-insulator transition (MIT) material. This wire is in contact on its two extremities with two themal reservoirs at temperature $T_L$ and $T_R<T_L$, respectively. The MIT material  is able to change its thermal conductivity following a hystereris curve (see Fig. 1-b) with respect to the temperature around a critical temperature $T_c=341K$. Beyond $T_c$ the wire tends to become metallic (amorphous) and for sufficiently high temperatures  its thermal conductivity is $\Lambda_{m}\approx 6 W.m^{-1}.K^{-1}$. At the opposite, below $T_c$ $VO_2$ tends to be crystalline (i.e. insulating) and $\Lambda_{d}\approx 3.6 W.m^{-1}.K^{-1}$ at sufficiently low temperature. However the evolution between these two extremes values follows a hysterezis loop~\cite{Oh} with respect to the temperature. Recent works have demonstrated that the thermal  bistability of these MITs can be exploited to store thermal information~\cite{BaowenLi3,Xie,Slava,DyakovMemory}.

\begin{figure}
\includegraphics[angle=0,scale=0.26]{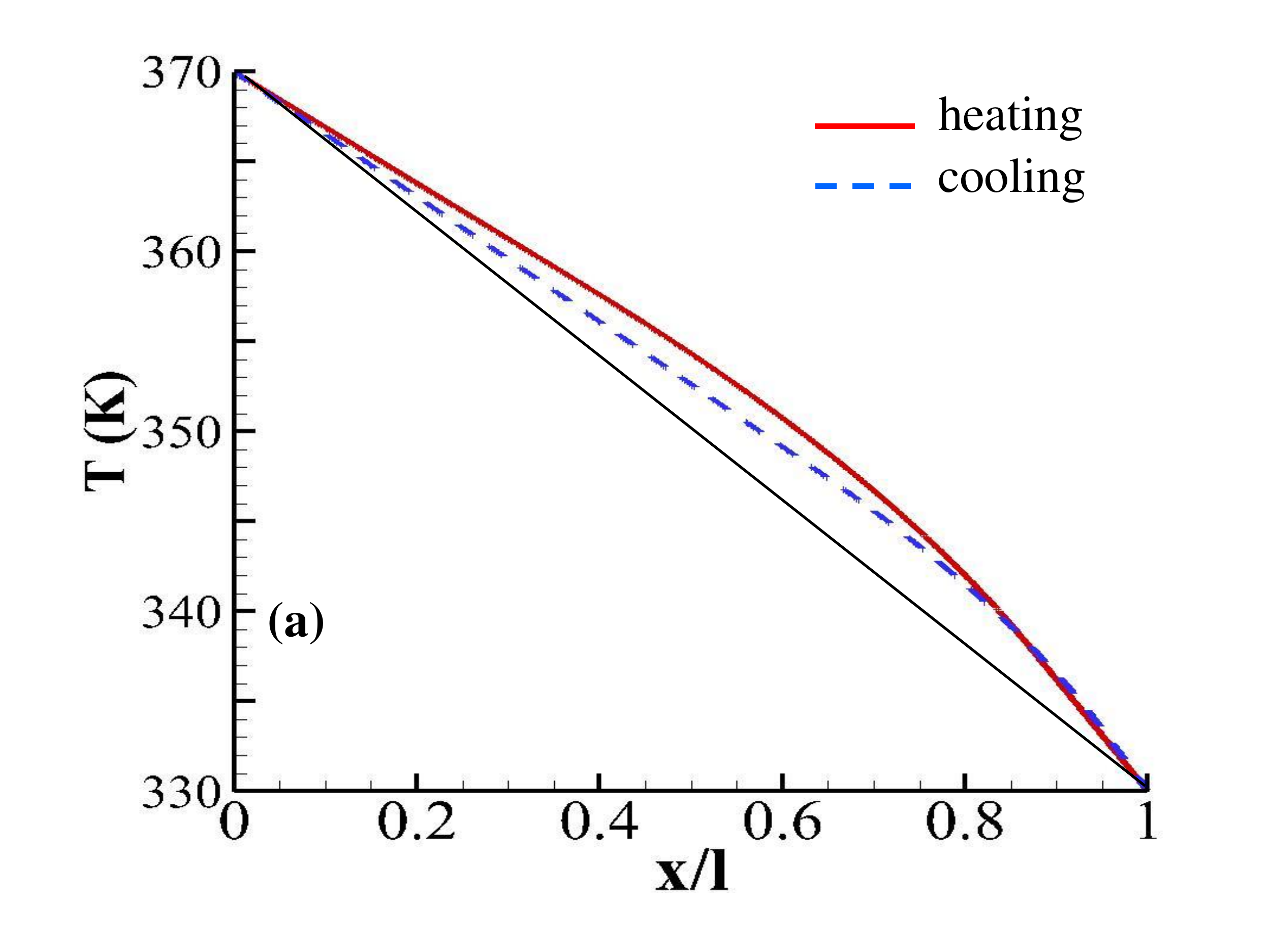}
\includegraphics[angle=0,scale=0.25]{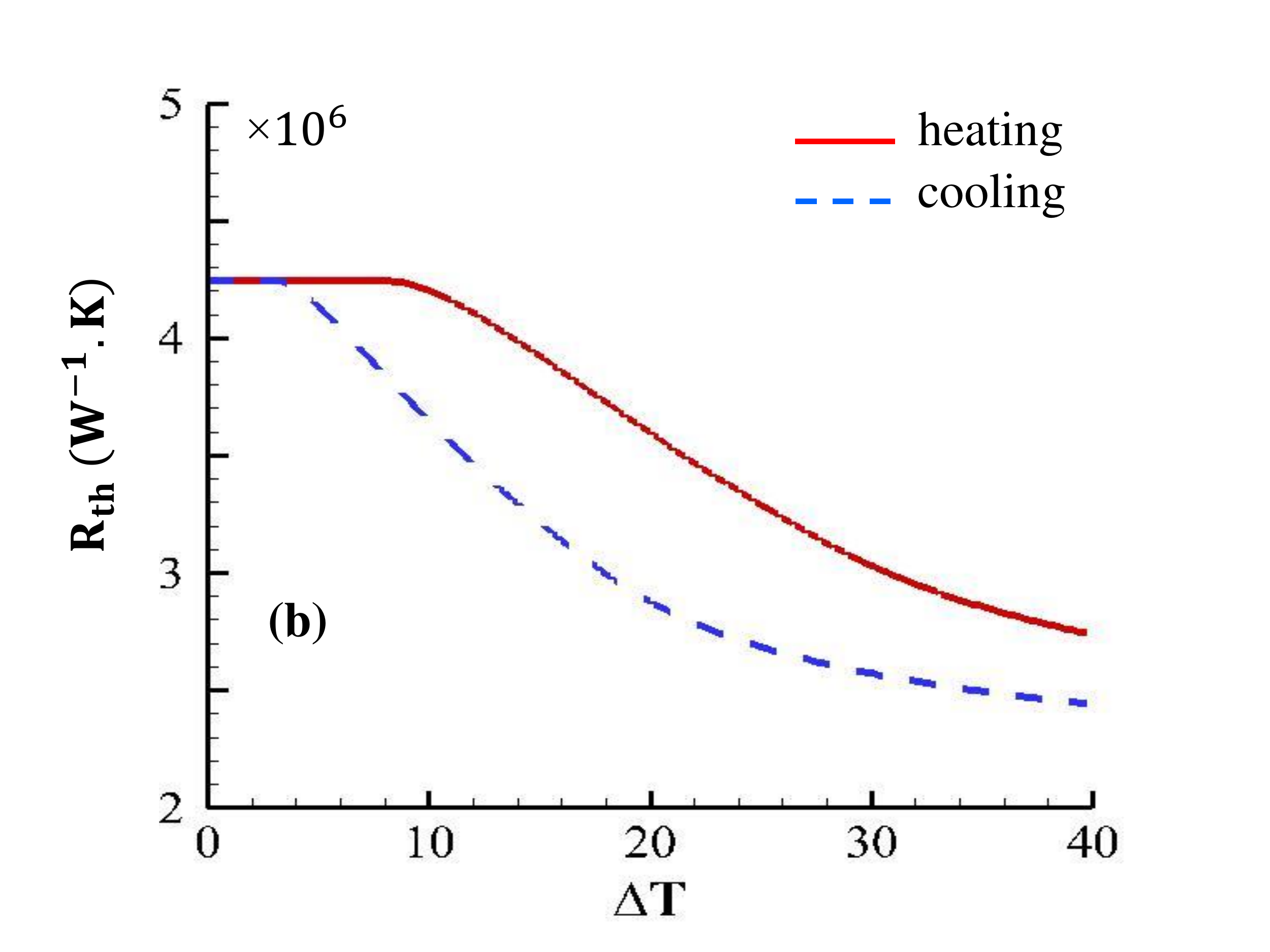}
\includegraphics[angle=0,scale=0.25]{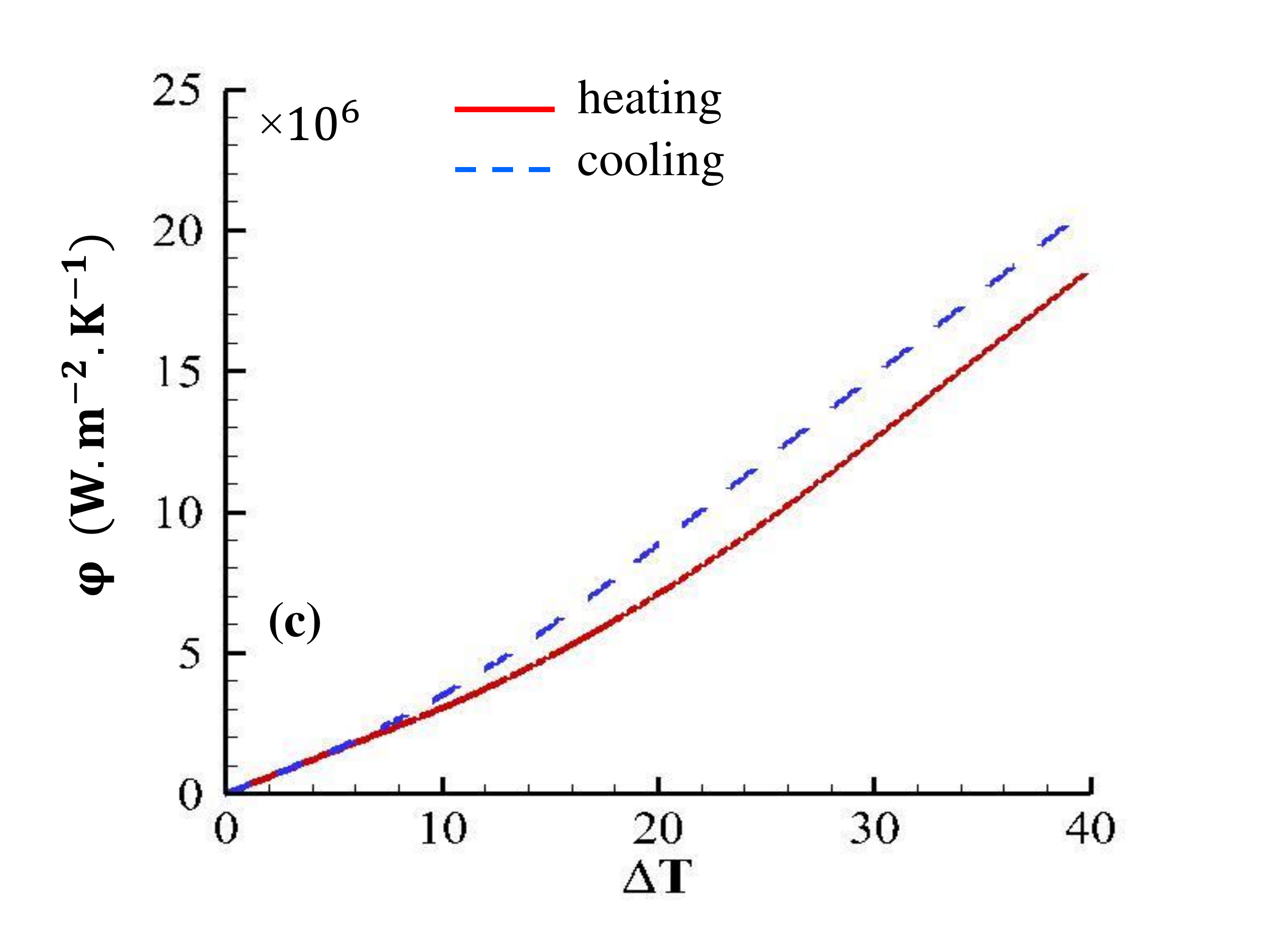}
\caption{(a) Temperature profile along a $VO_2$ thermal memristor $10 \mu m$ length and $500 nm$ radius when $T_L=370 K$ while $T_R=330 K$. The lower curve (solid black line) corresponds to the temperature profile in the diffusive approximation. (b) Thermal resistance with respect to the temperature difference $\Delta T=T_L-T_R$ when $T_R=330 K$ and  $T_L>T_R$. (c) Heat flux crossing the wire with respect to the temperature difference when $T_R=330 K$.}
\label{Fig_2}
\end{figure}

By changing the temperature gradient along the wire, the phase front moves along the wire so that its thermal resistance $R_{th}$ changes with respect to the temperature difference $\Delta T=T_L-T_R$ between the two reservoirs. In steady state regime and without convection on the external surface of wire, the heat conduction obeys the following equation
\begin{equation}
\frac{d}{dx}(\Lambda(T(x))\frac{dT(x)}{dx})=0. \label{Eq:conduction}
\end{equation}
By applying a Kirchoff's transformation 
\begin{equation}
W(x)=\int_{T_R}^{T(x)}\Lambda(\tilde{T}(x)) d\tilde{T}, \label{Eq:Kirchoff}
\end{equation}
on the thermal conductivity $\Lambda$, it is straighforward to show that the temperature profile along the wire is solution of the following equation
\begin{equation}
\int_{T_R}^{T(x)}\Lambda(\tilde{T}(x)) d\tilde{T}= \int_{T_R}^{T_L}\Lambda(\tilde{T}(x)) d\tilde{T}(1-\frac{x}{l}).\label{Eq:temperature1}
\end{equation}
Using the piecewise decomposition
\begin{equation}
\Lambda(T)= \left\{
\begin{array}{c}
\Lambda_d,\quad\quad\quad\quad\quad\quad\quad\quad\ T<T_1^{(i)}\\ 
a_i T+b_i,\quad\quad\quad T_1^{(i)}<T< T_2^{(i)}\\
\Lambda_m,\quad\quad\quad\quad\quad\quad\quad\quad T>T_2^{(i)}
\end{array}
\right.
\end{equation}
 of the conductivity with respect to the temperature, an explicit expression for the temperature profile can be derived \cite{SupplMat} from (\ref{Eq:temperature1}). In this decomposition, the subscript $i$ refers to the heating ($i=1$) or the cooling ($i=2$) phase.

The flux $\varphi$ flowing accross a wire of section $S$ is related to the temperature difference $\Delta T$ between the two reservoirs and to the thermal resistance $R_{th}$ by the simple system of equations
\begin{equation}
\begin{split}
\Delta T=R_{th}(x_1^{(i)},x_2^{(i)},\Delta T)S\varphi,
\\ \frac{dx_j^{(i)}}{dt}=g^{(j)}(\Delta T), \quad  j=1,2 \label{Eq:memristor}
\end{split}
\end{equation}
where the state variables $x_1^{(i)}$ and $x_2^{(i)}$ represent here the location along the wire below (resp. beyond)  which the MIT material becomes metallic (resp. insulating) during the heating or cooling  phase. $g^{(j)}$ denotes the function which drive the location of phase front inside the wire. Note that, depending on the thermal boundary conditions applied on the wire, these points can potentially be located outside of the wire (see \cite{SupplMat}). The thermal resistance  is obtained by summing the resistances in series $R_1^{(i)}$ along wire below $x_2^{(i)}$, $R_2^{(i)}$ between $x_2^{(i)}$ and  $x_1^{(i)}$ and  $R_3^{(i)}$ beyond $x_1^{(i)}$. Accordingly 
\begin{equation}
R_{th}(x_1^{(i)},x_2^{(i)},\Delta T)=\underset{j}{\sum}R_j^{(i)} \label{Eq:resistance}
\end{equation}
with $R_1^{(i)} =\frac{x_2^{(i)}}{\pi r^2\Lambda_m}$, $R_2^{(i)} =\frac{1}{\pi r^2}\int_{max(x_2^{(i)},0)}^{min(x_1^{(i)},l)}\frac{dx}{a_i T(x)+b_i}$ and $R_3^{(i)} =\frac{l-x_1^{(i)}}{\pi r^2\Lambda_d}$. Notice that these resistances can vanish depending on the value of state variables $x_1^{(i)}$ and $x_2^{(i)}$.
\begin{figure}
\includegraphics[angle=0,scale=0.25]{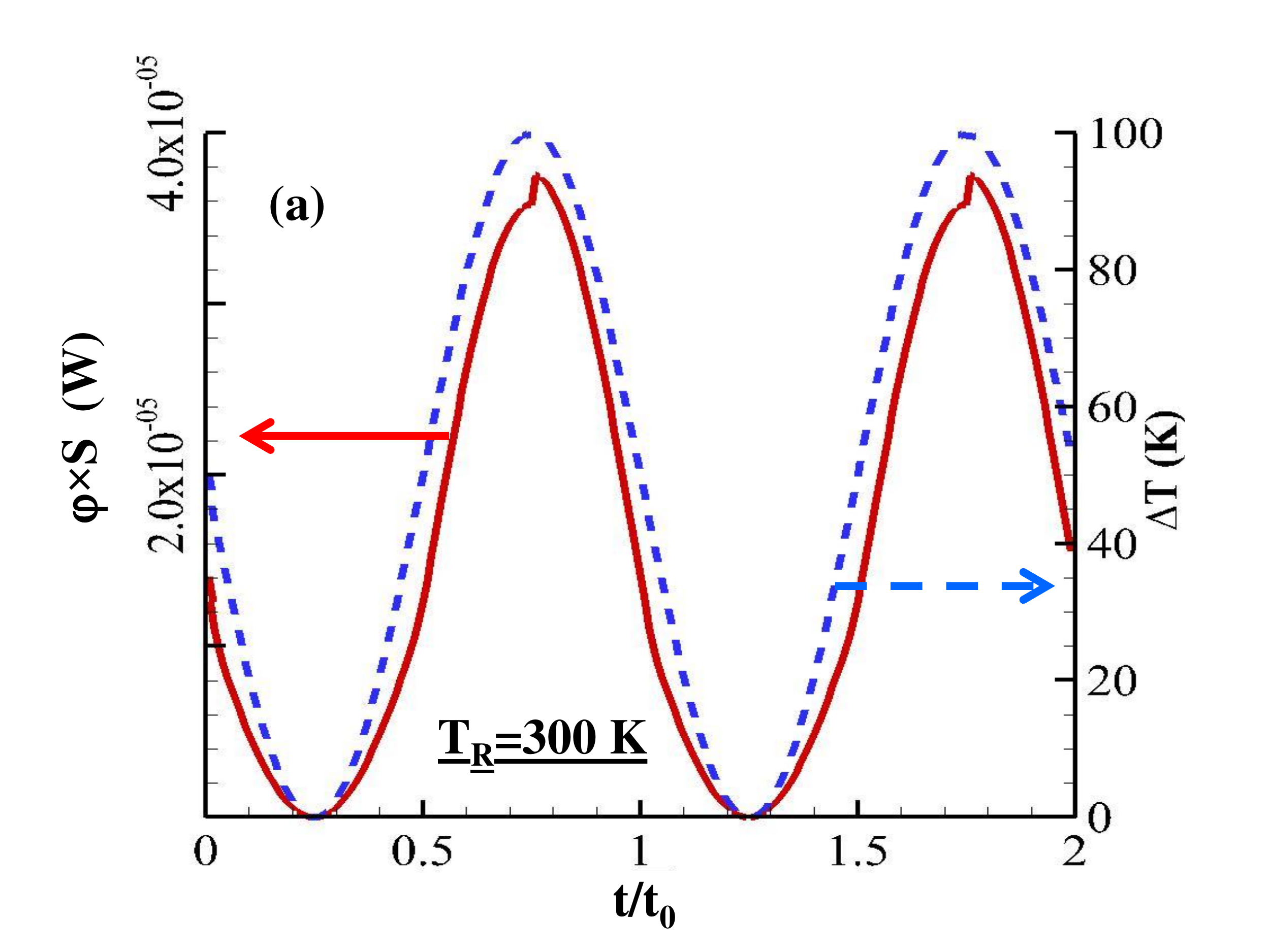}
\includegraphics[angle=0,scale=0.25]{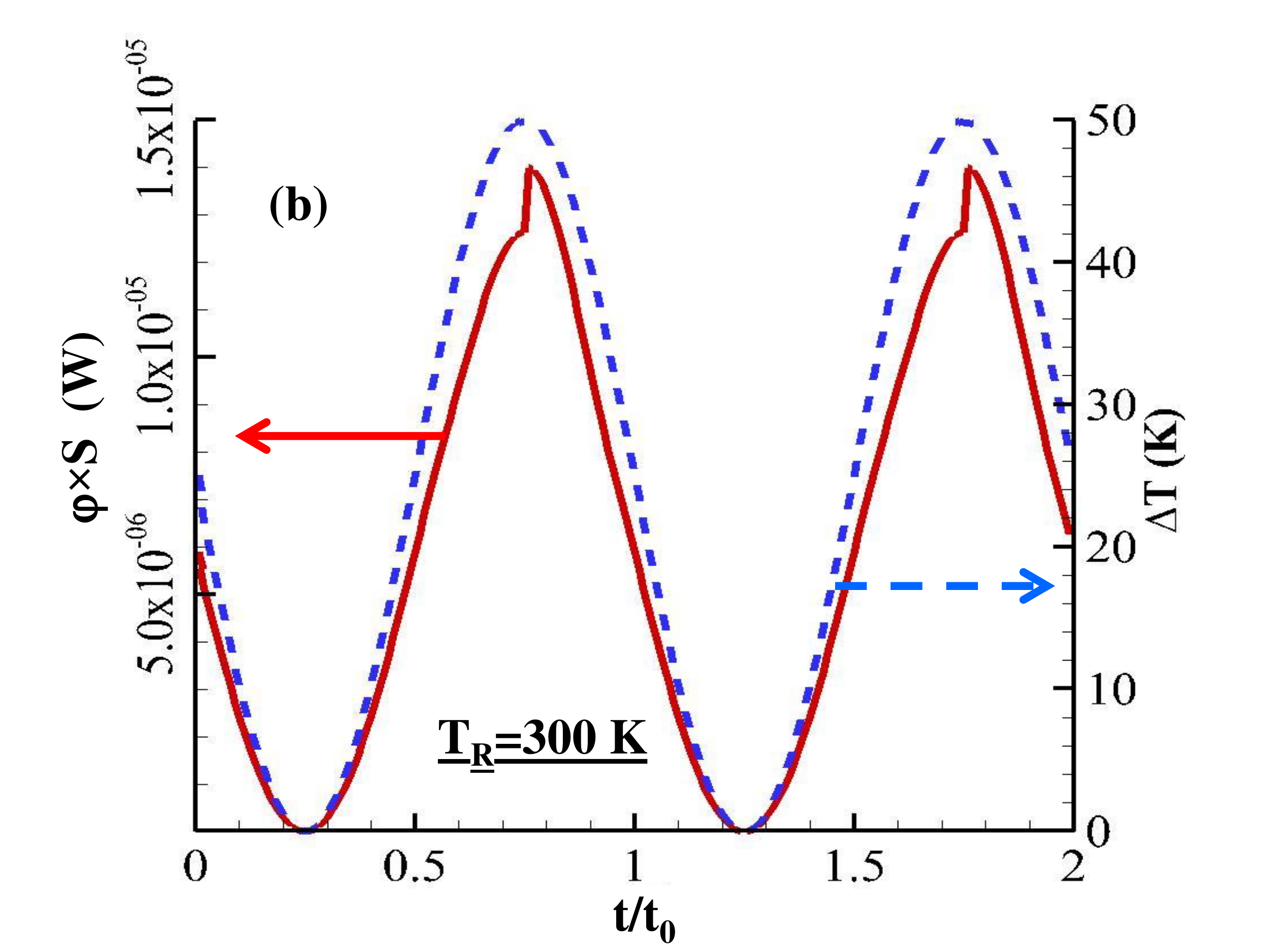}
\includegraphics[angle=0,scale=0.25]{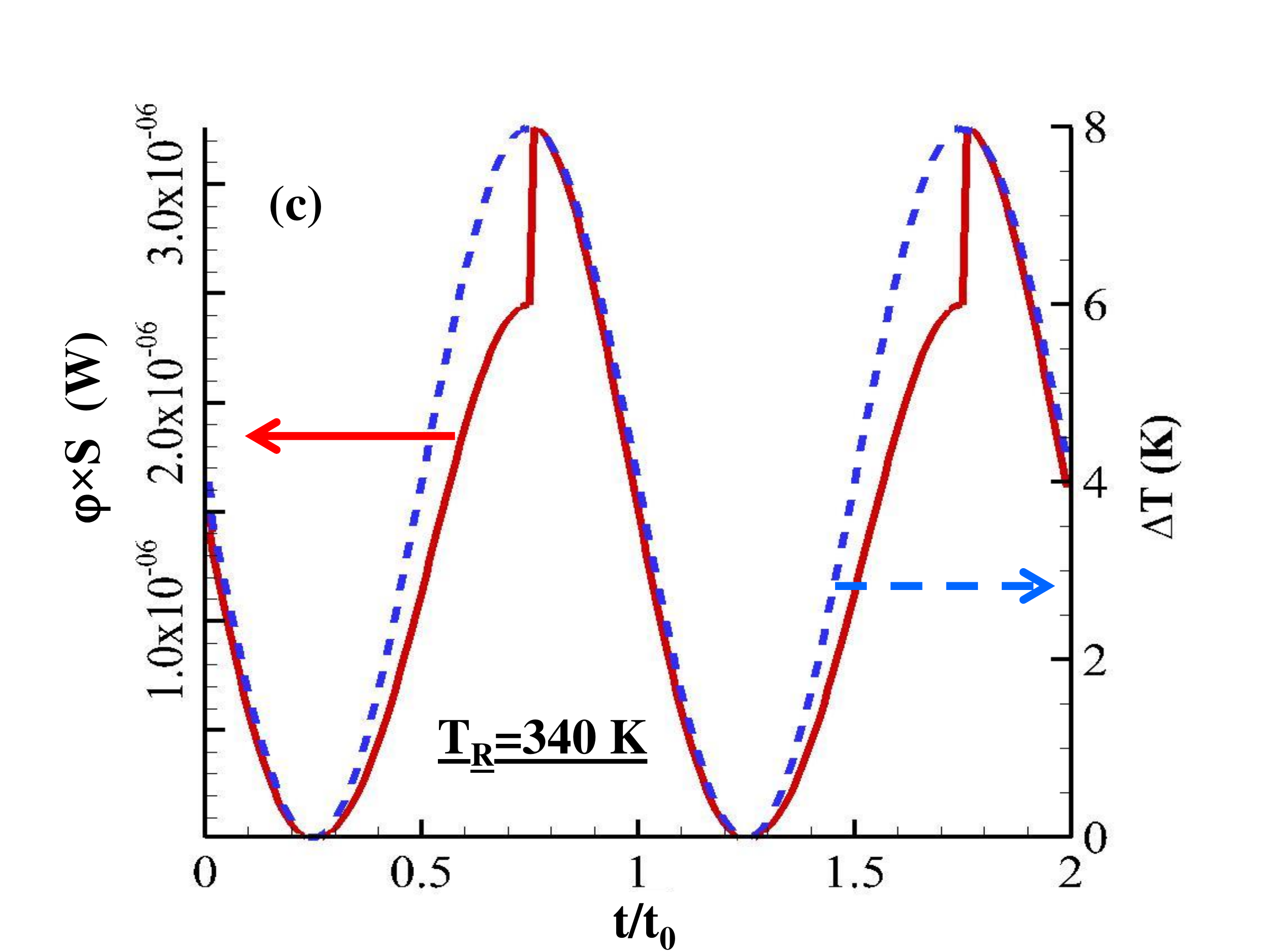}
\caption{ Power flowing through a $VO_2$ memristor $500\ nm$ radius and $10 \mu m$ length under an applied sinusoidal temperature difference $\Delta T(t)=T_L(t)-T_R$  for different temperature $T_R$. $t_0$ denotes the period of oscillations of temperature differences.}
\label{Fig_3}
\end{figure}
In Fig. 2(a) are plotted the temperature profiles along a $VO_2$ wire under a temperature gradient during the cooling and heating steps. Contrary to a classical conduction process those temperature profiles are not linear because of the temperature dependence of the thermal resistance as shown in Fig. 2(b). It results from this nonlinearity a nonlinear variation of flux crossing the wire as well (Fig. 2(c)) with respect to the temperature difference applied on it. 

Following a similar approach as in the Chua${'}$~s work~\cite{Chua}, based on symmetry arguments, we can derive the equivalence rules  between the electronic and the thermal problem for the fundamental quantities.
\begin{equation}
\begin{split}
\varphi \leftrightarrow i,
\\ T \leftrightarrow v,
\\ R_{th}=\frac{dT}{d\varphi}\leftrightarrow R,
\\q_{th}=\int\varphi dt\leftrightarrow q, \label{Eq rules}
\end{split}
\end{equation}
where $q_{th}$ and $q$ denote the thermal charge obtained by time integration of heat flux and the classical electric charge, respectively. It follows that the thermal capacity can be defined as
\begin{equation}
C_{th}=\frac{dq_{th}}{dT}=\varphi (\frac{dT}{dt})^{-1}. \label{capacity}
\end{equation}
As for the thermal analog of magnetic flux, it is given by
\begin{equation}
\Phi_{th}=\int T dt. \label{magnetic flux}
\end{equation}
Finally, the thermal memristance ("the missing element") reads
\begin{equation}
M_{th}=\frac{d\Phi_{th}}{dq_{th}}=\frac{d\Phi_{th}}{dt}(\frac{dq_{th}}{dt})^{-1}=\frac{\varphi}{T}.\label{memristance}
\end{equation}

A key point  for a memristor is its operating mode under a transient excitation.  Provide the timescale at which the boundary conditions varie is large enough compared with the relaxation time of temperature field inside the wire itself, the variation of flux crossing the system can be calculated from relation (\ref{Eq:memristor}). In typical solids, the thermalization timescale varies between few picosecond at nanoscale (phonon-relaxation time) to few microsecond at microscale (diffusion time $t\sim l^2/\alpha$, $\alpha$ being the thermal diffusivity).  The application of an external bias $\Delta T(t)$ across the system moves the position of state variables  $x_1^{(i)}$ and $x_2^{(i)}$ causing a time evolution of the thermal resistance $R_{th}$.  In Fig. 3 we show the time evolution of this flux for a sinusoidal variation of $\Delta T(t)$ with a period $t_0$ large enough compared to the relaxation time of memristor. Since the effective conductance (resp. resistance) of memristor increases (resp. decreases) when the system switch from the heating to the cooling phase we observe a significant enhancement of heat flux (up to $25\%$) flowing through the memristor each time $T_L$  decays. This sharp variation of physical properties can be exploited to design basic neuronal circuits and make logical operations with thermal signals.

\begin{figure}
\includegraphics[angle=0,scale=0.4]{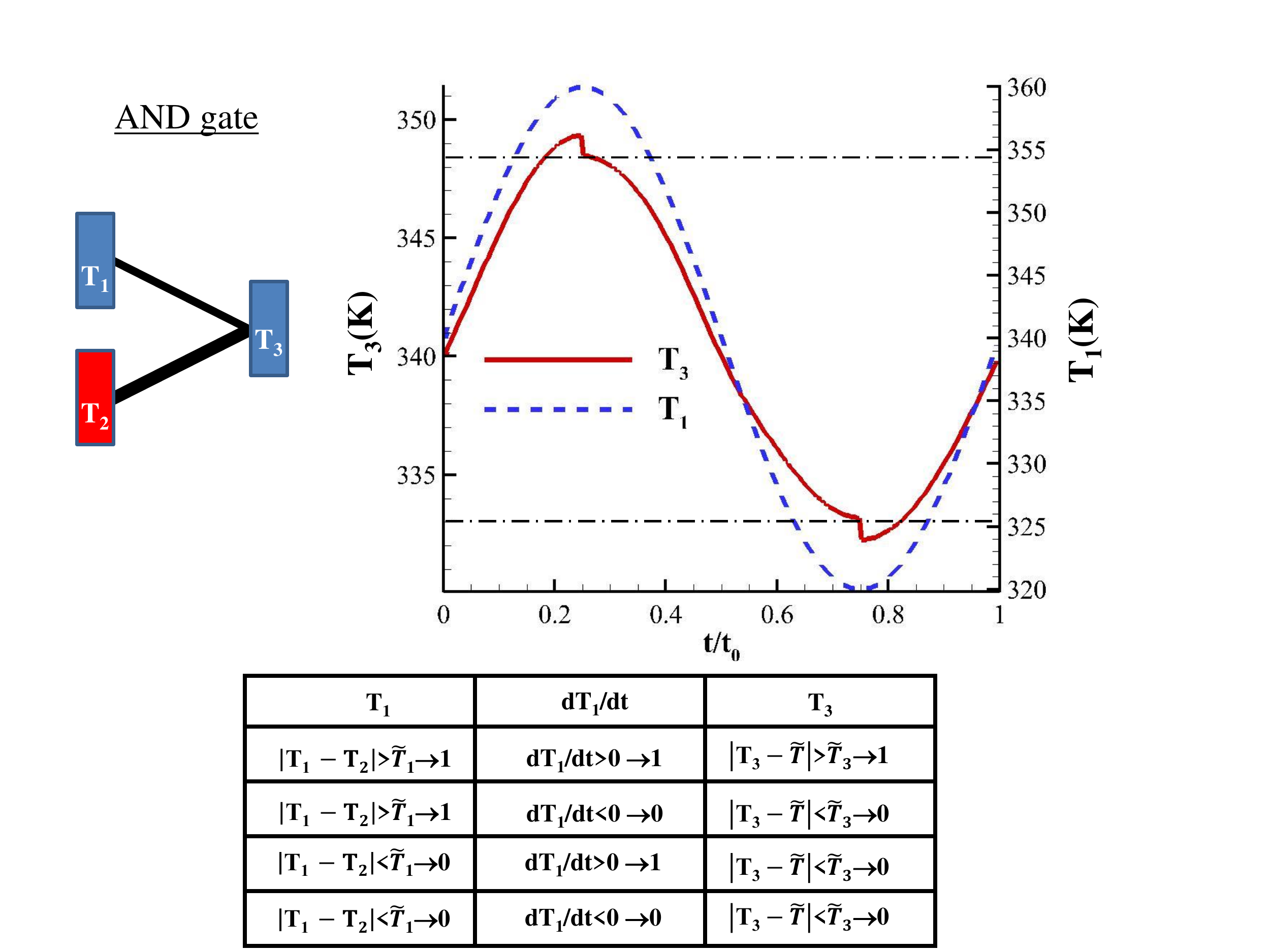}
\caption{Neuronal AND gate made with two phase-change memristors with one fixed node at $T_2=340 K$. The two inputs signals are $T_1$ and its derivative $dT_1/dt$ with respect to time and the gate truth table is summarized on the array. The  two treshold temperatures $\tilde{T}_{1}$ and  $\tilde{T}_{3}$ (dashed horizontal lines) correspond to the temperatures where the system switches from its heating (resp. cooling) to its cooling (resp. heating) operating mode. Here, the two meristors have the same length ($10 \mu m$) and two different radius ($400\ nm$ and $500\ nm$).}
\label{Fig_4}
\end{figure}

Recent works have demonstrated the possibility to make such logical operations with acoustic phonons~\cite{BaowenLi2,BaowenLi3,BaowenLiEtAl2012} or thermal photons~\cite{OteyEtAl2010,BasuFrancoeur2011,Huang,Dames,PBA_APL,Ito,van Zwol1,PBA_PRL2014} by using phononic and photonic counterpart of diodes and transistors. 
Here, we demonstrate that memristive systems can be an alternative to these systems. To show that, let us consider a simple neuron~\cite{McCulloch} made with two memristors connected to the same node (output) as sketched in Fig. 4. One temperature ($T_1$) is used as an input signal while the second temperature $T_2$ plays the role of a simple bias and is held at a fixed value. The time variation of the first input (related to the power added or extracted from the solid in contact to the left side of first memristor) set the second input parameter. Depending on its sign (i.e. heating or cooling process) this parameter can be assimilated to a binary parameter as shown on the truth table in Fig. 4.
Then, the output temperature $T_3$ is obtained with respect to $T_1$ and $T_2$ by solving,  in steady state regime,  the energy balance equation
\begin{equation}
\underset{j\neq3}{\sum} R_{th}^{-1}(T_3,T_j)(T_j-T_3)=0. \label{Eq:temp_and}
\end{equation}
According to the change in the two inputs signals, a sharp transition for the output can be observed (Fig. 4). This transition occurs precisely when the memristors switches from their heating to their cooling operating mode. 
The horizontal lines in Fig.4 given by $\left|T_1-T_2\right|=\widetilde{T_1}$ and $\left|T_3-\widetilde{T}\right|=\widetilde{T_3}$  (where $\widetilde{T}=(T_{3,min}+T_{3,max})/2$) allow to define the $0$ and $1$ states for the gate. By using the truth table shown in Fig. 4 it appears that this simple neuron operates as an AND gate. Beyond this  logical operation more complex neuronal architectures, where the memristors  are used as  on/off temperature dependent bistable switchs,  can be designed to implement other boolean operations.


To summarize, we have introduced the concept of  phase-change thermal memristor and shown that it constitutes a fundamental  building block for the implementation of basic logical operations with neurons entirely driven by heat. The relaxation dynamic of these memristors combined with the massively parallelism of neuronal networks make promising these systems both for thermal computing and active thermal management at a submicronic time scale. 

\section*{ Supplementary Material}
In the supplementary material we give the analytical expression of $\Gamma$ parameter and of temperature field inside a memristor with respect to the temperature difference apply on it.

\section*{ Acknowledgments}
P.B.-A. acknowledges discussions with Dr I. Latella

\end{document}